\documentclass[preprint]{aastex}
\usepackage{psfig}
\slugcomment{Accepted by AJ}
\shorttitle{Merging Pairs in the SDSS EDR}
\shortauthors{Allam et al.}
\begin{document}

\title{Merging galaxies in the SDSS EDR}

\author{
Sahar S.\ Allam\altaffilmark{1,2},
Douglas L.\ Tucker\altaffilmark{2},
J.\ Allyn Smith\altaffilmark{3,4},
Brian C.\ Lee\altaffilmark{5},
James Annis\altaffilmark{2},
Huan Lin\altaffilmark{2},
Igor Karachentsev\altaffilmark{6},
Bryan E.\ Laubscher\altaffilmark{3}}
\altaffiltext{1}{Dept.\ of Astronomy, New Mexico State University, Las Cruces, NM 88001}
\altaffiltext{2}{Fermi National Accelerator Laboratory, P.O. Box 500, Batavia, IL 60510}
\altaffiltext{3}{Los Alamos National Laboratory, ISR-4, MS D448, Los Alamos, NM 87545}
\altaffiltext{4}{University of Wyoming, Dept.\ of Physics \& Astronomy, P.O. Box 3905, Laramie, WY 82071}
\altaffiltext{5}{Lawrence Berkeley National Laboratory, 1 Cyclotron Rd, Berkeley CA 94720-8160}
\altaffiltext{6}{Special Astrophysical Observatory, Russian Academy of Sciences Zelenchukskaya, 357147, Russia} 

\begin{abstract}
We present  a  new catalog  of  merging  galaxies obtained through  an
automated systematic search routine.   The  1479 new pairs of  merging
galaxies were  found in $\approx$ 462~sq~deg of  the Sloan Digital Sky
Survey   Early    Data   Release  (SDSS   EDR;  \citealt{Stoughton02})
photometric data, and the pair catalog is complete for galaxies in the
magnitude range $16.0 \leq g^* \leq 20$.

The selection algorithm, implementing a variation on the original
\citet{Karachentsev72} criteria, proved to be very efficient and fast.
Merging galaxies were  selected such that the inter-galaxy separations
were less than the sum of the component galaxies' radii.

We discuss the characteristics of the sample in terms of completeness,
pair  separation, and the Holmberg  effect.  We also present an online
atlas of  images for the SDSS EDR  pairs obtained  using the corrected
frames from the SDSS EDR database.   The atlas images also include the
relevant data for each pair member.

This catalog  will  be useful for  conducting studies   of the general
characteristics of  merging  galaxies, their  environments,  and their
component galaxies.  The redshifts for a subset of the interacting and
merging galaxies and   the  distribution of angular  sizes   for these
systems indicate the SDSS provides  a  much deeper sample than  almost
any other wide-area catalog to date.
\end{abstract}

\section{Introduction}

Interacting  and  merging  galaxies are  among   the most  fascinating
astronomical  objects  in the Universe.   These galaxy  systems span a
wide range of configurations, from single  distant encounters to close
encounters  which  may  result in  a  single  merged system.   Typical
morphologies   include bridges between    the interacting partners and
tidal  tails.  These structures are usually  associated  with sites of
strong star  formation resulting in dense star  clusters or even dwarf
galaxy sized  objects.   Additionally, a  central  star burst might be
triggered by tidally induced bars which  funnel matter to the galactic
center.  This  coupling between galactic   dynamics and star formation
processes provides    a unique  tool  for  a  deeper understanding  of
galactic evolution.

From a theoretical standpoint, interactions and mergers are considered
to be major drivers of galactic evolution, affecting morphological and
spectral  characteristics, gas-dynamics,    high-energy processes, and
nuclear activity.  A  merger sequence for  the formation of elliptical
galaxies   from    interacting    spirals   was first     proposed  by
\citet{Toomre77} based  on  numerical  simulations.  Inclusion of  gas
dynamics and star formation  using Tree-SPH codes  is now the state of
the art  (cf.  \citealt{MihosHernquist96})  and allows  a   variety of
predictions for the induced star  formation and dynamical phenomena to
be compared with observations.  A review of these processes and of the
growing   evidence  that mergers  play a    major  role in the delayed
formation of elliptical and early-type disk galaxies both in the field
and in clusters can be found in \citet{Schweizer99}.

Observations of galaxy pairs at intermediate redshifts have revealed a
larger number    of  such systems in the     past (\citealt{ZepfKoo89,
Burkey94, Carlberg94, Yee95, Patton97, LeFevre00}, but see also
\citealt{Woods95}), confirming that merging plays an important role in
galaxy evolution.  This is suggested,   for example, by the excess  of
blue   star-forming        galaxies     at   intermediate    redshifts
\citep{ButcherOemler84}.  However, most previous work has been limited
to small  samples of objects  covering a  small area on  the sky, thus
making  the statistical  analyses less   than ideal.

The  best known catalogs of interacting  and merging pairs of galaxies
are, course, those by Vorontsov-Velyaminov \citeyearpar{VV59,VV77} and
\citet{Karachentsev72}, and  many  interacting and  merging  pairs can
also  be found in  \citet{Arp66}'s {\em  Atlas  of Peculiar Galaxies}.
The Vorontsov-Velyaminov catalog of interacting galaxies, updated in a
web edition\footnote{{\tt http://www.sai.msu.su/sn/vv/}} in 1999 by R.
I.   Noskova   \&  V.   P.   Arkhipova of   the Sternberg Astronomical
Insitute, contains 2014 systems, mostly confined to declinations north
of $-45^{\circ}$,  complete for galaxies brighter   than $m_{\rm pg} =
15$.  The Vorontsov-Velyaminov  catalog is not restricted to  isolated
galaxy pairs,  but also  contains    peculiar single galaxies,   tight
groups, ``nests'' \citep{Zasov00}, and chains of galaxies.

\citet{Karachentsev72}, on   the  other hand,  searched explicitly for
galaxy pairs  that   were  isolated.    Using  \cite{Zwicky68}'s  {\em
Catalogue  of Galaxies \& Clusters  of  Galaxies\/} as his source  for
galaxy  positions and  magnitudes, he   identified 603 isolated  pairs
north of $\delta_{\rm 1950} = -3^{\circ}$ for galaxies down to $m_{\rm
pg} = 15.7$.   Due to its  size, completeness, and relatively unbiased
selection,  this  catalog has  been a popular    sample for studies of
isolated    galaxy   pairs  (\citealt{Oleak76, Stocke78,  Arakelian81,
Karachentsev81,  Picchio85,   Sulentic89,   Keel91,  Petrov92, Flin93,
Sulentic92, Keel96,   Allam98, Toledo99,    Byrd01,      Teerikorpi01,
Franco-Balderas03}; among others).

Other   catalogs of  galaxy   pairs   include   an extension  of   the
Karachentsev  catalog by  \citet{ReduzziRampazzo95}, which  netted 409
candidate isolated pairs, 214 of which lie in the Southern Hemisphere,
and  a   catalog  of  621    southern isolated pairs   identified   by
\citet{Soares95}  using a surface density   enhancement method.   Both
these  catalogs  were extacted  from the  {\em  The Surface Photometry
Catalogue of  the ESO-Uppsala Galaxies\/} \citep{Lauberts89}, which is
a galaxy catalog  complete    to $B \approx    14.5$ \citep{Soares95}.
Finally, the advent of large redshift surveys has  made it possible to
hunt   down   large,   homogeneous   samples  of  galaxy     pairs  in
velocity-space.  Important   among   these    are the  samples      by
\citet{Barton00}, who  found  251  close  pairs in   the CfA2 Redshift
Survey, and  by \citet{Lambas02}, who  found  1258 pairs  in the 100~K
public release of the 2dF Galaxy Survey.

Here, we present a catalog of 1479 merging pairs of galaxies extracted
from  the approximately  462~sq~deg  of imaging  data  from the  Sloan
Digital    Sky    Survey     Early    Data    Release    (SDSS    EDR;
\citealt{Stoughton02}).   We provide an  overview of  the SDSS  EDR in
\S~\ref{sec:DATA},     present    our     selection     criteria    in
\S~\ref{sec:SelectionCriteria},  describe   the  construction  of  the
catalog  in \S~\ref{sec:construction}, present  the catalog  itself in
\S~\ref{sec:Prop}, and draw conclusions in \S~\ref{sec:conclusion}.

\section{The Data} \label{sec:DATA}

The SDSS is a digital photometric and spectroscopic survey which will,
when  completed, cover  one quarter of  the   celestial sphere  in the
Northern Galactic Hemisphere and   an additional 225 square degree  in
the               Southern          Galactic                Hemisphere
\citep{York00,Eisenstein01,Hogg01,Smith02,Strauss02,Pier03}.       The
photometric mosaic  camera \citep{Gunn98} images  the sky  by scanning
along  great  circles  at the sidereal  rate.    The imaging  data are
produced simultaneously  in five  photometric bands ($u^{*}$, $g^{*}$,
$r^{*}$,   $i^{*}$,   and  $z^{*}$,    with effective   wavelengths of
$\lambda$3500,     $\lambda$4770,     $\lambda$6231,    $\lambda$7625,
$\lambda$9134 $\AA$, respectively; see \citealt{Fukugita96}).

In  June 2001,  the  SDSS  EDR  were   presented to the   astronomical
community.  The SDSS  EDR consists  of 462  square degrees of  imaging
data -- mostly along the celestial equator --  in five bands, together
with  medium resolution spectra for  approximately 40,000 galaxies and
4,000 quasars.

We have downloaded a catalog of objects classified  as galaxies by the
SDSS imaging reduction software ({\tt photo}, see
\citealt{Lupton01,Lupton02}) from the SDSS   public archives and  used
this as  our  base catalog from   whence we extracted this catalog  of
galaxy pairs (see \S~\ref{sec:construction}).

\section{Selection Criteria}\label{sec:SelectionCriteria}

We have developed  a  systematic search criterion for  interacting and
merging galaxies  that can be directly  applied to large  data sets in
the public domain ---  such as the SDSS EDR.    We have made use of  a
method  developed  by \citet{Karachentsev72} to  search  for candidate
pairs.   Two  galaxies of  angular diameter  $a_{1}$ and $a_{2}$ whose
separation is $x_{1,2}$, will satisfy the Karachentsev criterion if
\vspace{-.2cm}
\begin{eqnarray*}
\frac{x_{1,i}}{x_{1,2}} & \ge  & \chi \frac{a_{i}}{a_{1}} , \\
\frac{x_{2,i}}{x_{1,2}} & \ge  & \chi \frac{a_{i}}{a_{2}} ,
\end{eqnarray*}
where $i$ indicates each of the  neighbor galaxies whose 
diameter $a_{i}$ is included in the intervals set by
\vspace{-.2cm}
\begin{center}
 $\xi a_{1}  \leq a_{i} \leq  \lambda a_{1}$ , \\
 $\xi a_{2}  \leq a_{i} \leq  \lambda a_{2}$ ,
\end{center}
\vspace{-.2cm}
where $\xi = 1/2$, $\chi = 5$, and $\lambda = 4$. 

\citet{Toledo99} found that the  Karachentsev pair sample  was  biased
toward binaries with a small viewing angle  and perhaps with a certain
orbital phase.  This  bias  could be  due   to the use of   the galaxy
diameter   criterion.   Spiral galaxies   have  complex   disks  whose
diameters     can   be   either   systematically   underestimated   or
systematically overestimated.  Elliptical galaxies, on the other hand,
show de~Vaucouleurs ($R^{1/4}$) profiles, and  have no clearly defined
``edge'';  measurements  of the diameters   of elliptical galaxies are
thus often underestimated.

To overcome this problem we have used  the Petrosian radius (Petrosian
1976).  The  Petrosian  radius is a metric  radius;  i.e.  its angular
dimension is given  by a formula  relating the physical dimension of a
rigid rod and its angular dimension.   For the SDSS EDR, the Petrosian
radius is defined as the largest radius within which the local surface
brightness is at least one-fifth  the mean surface brightness interior
to that radius (see, e.g.,  \S~4.4.5.4 of \citealt{Stoughton02}).  For
objects with the  same shape, the luminosity  within a fixed Petrosian
radius gives a fixed fraction of  the total luminosity.  Moreover, the
Petrosian radius does not depend on quantities that usually affect the
surface brightness such  as galactic absorption, cosmological dimming,
k-correction, or luminosity evolution.

\section{Catalog Construction} \label{sec:construction}

We  selected for our  base catalog  all SDSS   EDR objects which  were
classified   as  galaxies by  {\tt  photo};   we imposed the  addition
requirements that these galaxies  have Petrosian radii $R_{petro} > 0$
in  all   5   SDSS bands,    and   errors  in   the Petrosian    radii
$\sigma_{R_{petro}} >  0$ in  the $g^{*}$ band.   The total  number of
SDSS EDR galaxies  which satisfy these criteria  is 6,168,836. For our
candidate  pairs, we  considered  galaxies with  $g^*$-band  Petrosian
magnitudes
$16.0 \leq g^* \leq 21.0$.

To obtain  isolated pairs,  we imposed the  following  criteria on any
neighboring galaxy $i$:
\begin{center}
 $ | g^{*}_{1,2} - g^{*}_{i} | >  3.0 $ \\
 $0.25 \times R_{petro,1} \leq R_{petro,i} \leq 4.0 \times R_{petro,1}$ 
($g^*$-band)\\
 $0.25 \times R_{petro,2} \leq R_{petro,i} \leq 4.0 \times R_{petro,2}$ 
($g^*$-band) 
\end{center}
\vspace{-.25cm}
Finally, we classified each resulting galaxy pair by the separation of
its two members:
\vspace{-.25cm}
\begin{description}
\item[merging pairs:] 
$x_{1,2} \leq (R_{petro,1} +  R_{petro,2})$ 
\vspace{-.25cm}
\item[intermediate pairs:]
$(R_{petro,1} +  R_{petro,2}) < x_{1,2} \leq 3 \times (R_{petro,1} + R_{petro,2})$
\vspace{-.25cm}
\item[wide pairs:]
$3 \times (R_{petro,1} + R_{petro,2}) < x_{1,2} \leq 10 \times (R_{petro,1} + R_{petro,2})$ 
\vspace{-.25cm}
\end{description}

Throughout the remainder of this paper, we will concern ourselves only
with the merging pair   sample.   Our initial sample   contained  3385
merging  pair candidates.  To remove spurious  pairs due to poor image
deblending          (\S~4.4.3         of        \citealt{Stoughton02};
\citealt{Lupton01,Lupton02}), one  of  us  (Allam)  inspected all  the
merging   pairs  by  eye.   She  also used    {\tt SExtractor} on  the
$g^{*}$-band SDSS  FITS images, and  classified the pairs according to
their component intensities and  isolation.  After all rejections  and
verifications,  the final number  of  candidate merging pairs left for
inclusion in this catalog was 1479. Of the  1906 which were discarded,
955 were  false detections due to  poor deblending of saturated bright
stars, 477 were  angularly  large galaxies erroneously  fragmented  by
deblending, and  474 were faint diffuse  objects in  the original SDSS
image processing.

\begin{figure}
\centering
\makebox[150mm]{\psfig{file=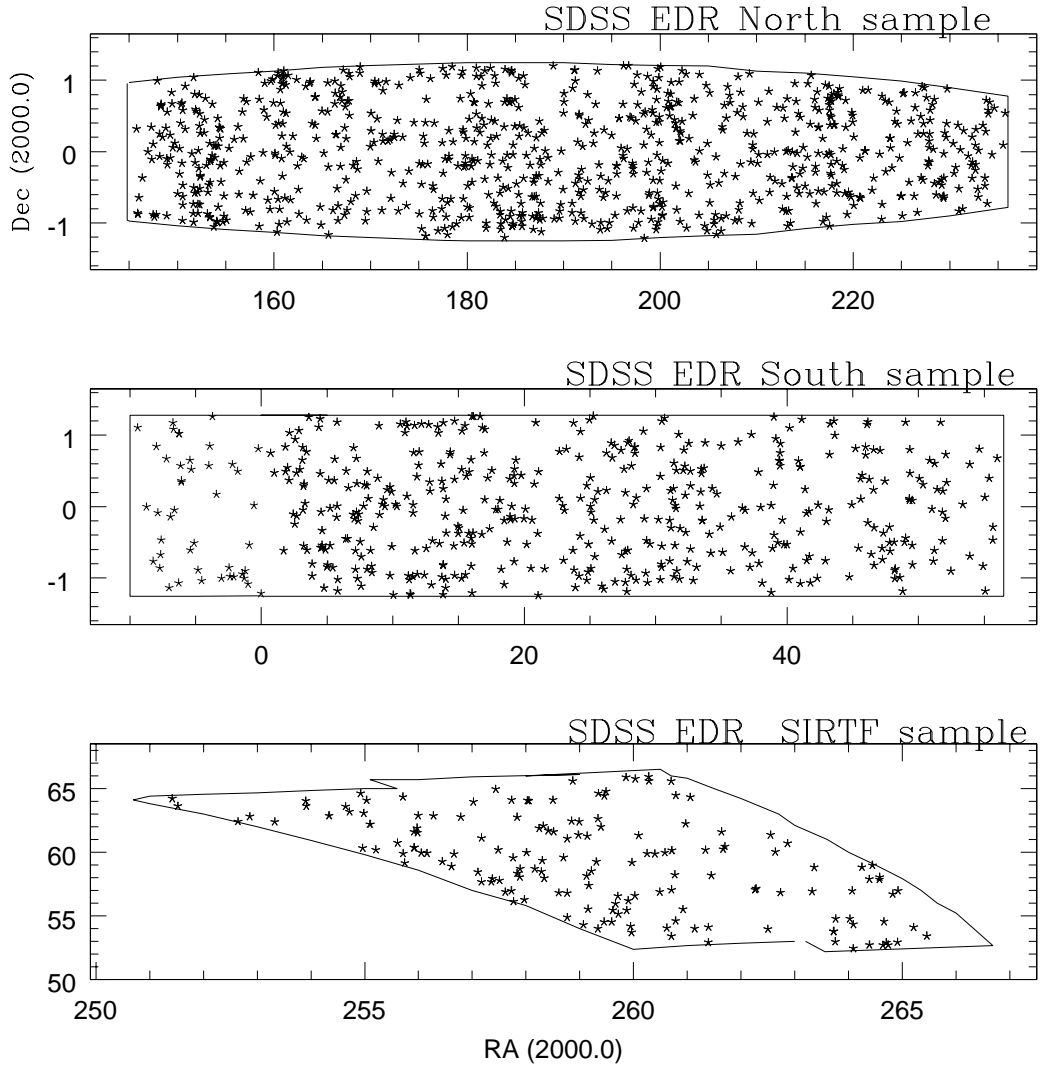,width=150mm,height=150mm,clip=,angle=0,silent=}}
\caption{The sky distribution of the 1479 merging pairs in RA/Dec 
(J2000.0)     for   the  three    different     areas  of    the  SDSS
EDR (the Northern equatorial sample, the Southern equatorial sample, 
and the SIRTF ``First Look'' fields).  \label{figradec}}
\end{figure}

Figure~\ref{figradec} shows  the sky distribution  of the 1479 merging
galaxy   pairs for the  three  different areas of  the   SDSS EDR (the
Northern equatorial  sample,  the Southern equatorial  sample, and the
SIRTF ``First  Look'' fields).  Table~\ref{tabinmerg}  lists, for  the
magnitude  range $g^*=16-21$, the total  number  of SDSS EDR  galaxies
($N_{\rm gal}^{\rm  tot}$),  the number of  galaxies  in merging pairs
($N_{\rm gal}^{\rm pair}$), the  mean surface density of  merging pair
galaxies on the sky ($\Sigma_{\rm  gal}^{\rm pair}$), and the fraction
of  all  galaxies that  are part of  merging pairs  ($f_{\rm gal}^{\rm
pair}$) for each SDSS  run in the SDSS EDR.   Note that about one-half
of  one percent of  all SDSS EDR galaxies are  in  merging pairs as we
have defined them.   The space distribution (redshift  vs.\ RA) of the
744 merging pair galaxies  with known spectroscopic redshifts is shown
in Figure~\ref{fig1}.

To estimate  the  number of  merging   pairs which are  merely  chance
projections in the final sample of 1479, we generated a random catalog
containing  245,924 galaxies at the same  mean  surface density on the
sky  as our real sample and   re-ran our algorithm.   Only one merging
pair was  found.  A  value  of $1\pm1$    spurious pairs  per  245,924
galaxies  (the  uncertainties   are based   on  $\sqrt{N}$ statistics)
implies $25\pm25$  spurious  pairs in our  base   catalog of 6,168,836
galaxies, yielding an estimated contamination by chance projections in
the final sample of  merging pairs of  less than 3.4\% ($25\pm25$  per 
1479 or 1.7\%$\pm$1.7\%).  (Since we    have  ignored  the  effects of
large-scale structure and the statistical clustering of galaxies, this
value   is   likely    an  underestimate.    As    we  will  show   in
\S~\ref{sec:Holmberg}, based   upon our limited  number  of pairs with
complete redshift information, the contamination rate may be more like
12.5\%.)

We note  that  \cite{Infante02} identified  pairs and small  groups of
galaxies in a subset of the SDSS EDR in  the magnitude range $18 < r^*
< 20$.  Our approach differs from theirs in  that we use the full SDSS
EDR sample, we use a different algorithm to identify pairs, and we use
{\tt  SExtractor} (version  2.0.15,  see \citealt{BertinArnouts96}) to
perform a final rigorous culling of our sample  to arrive at our final
catalog.

\section{Properties}\label{sec:Prop}

In Table~\ref{tabmerg} we list the general  properties of the SDSS EDR
pairs of merging galaxies.  The columns are as follows:

\noindent Column  (1): a running identification number.

\noindent Columns (2\&3): the pair's  RA and  Dec (J2000.0).

\noindent Columns (4--8): the pair's total SDSS Petrosian magnitude in 
each of the five bands ($u^{*}$, $g^{*}$, $r^{*}$, $i^{*}$, $z^{*}$).

\noindent Column (9): the projected separation between the two galaxies in arcsec.

\noindent Column (10): the SDSS name (SDSSJ~HH:MM:SS.ss$\pm$DD:MM:SS.s) for the 
brighter member of the pair in $g^*$ (galaxy ``a'').

\noindent Column (11): the spectroscopic redshift (if known) of galaxy ``a''.

\noindent Column (12): the SDSS name (SDSSJ~HH:MM:SS.ss$\pm$DD:MM:SS.s) for the 
fainter member of the pair in $g^*$ (galaxy ``b'').

\noindent Column (13): the spectroscopic redshift (if known) of galaxy ``b''.

The  basic global  properties of  this  sample are illustrated in  two
ways.  First,   the    mean  and median  properties  are    listed  in
Table~\ref{tabin}.  Second, the histograms of the intra-pair projected
separations, of the $u^*g^*r^*i^*z^*$ magnitudes of the pair galaxies,
and   of the {\em  differences}   of the  $u^*g^*r^*i^*z^*$ magnitudes
between   the     two  galaxies  in    each    pair  are  plotted   in
Figures~\ref{projsephist}, \ref{figh5bandmag}, and
\ref{fighindifmag}, respectively.

\subsection{The Atlas}
\label{sec:Atlas}

Due to the large number of pairs in  our catalog, we  do not provide a
hardcopy atlas.  Instead, we have prepared an  online atlas of merging
galaxy   pairs    from   the  SDSS   EDR    located   at   our  public
URL.\footnote{{\tt http://home.fnal.gov/$\sim$sallam/MergePair/}}  
The images were prepared from the $g^*$-band corrected frames obtained
from  the SDSS EDR  Data Archive   Server.  In  the atlas  images, the
center   is   marked by  a square   symbol  and the   pair members are
identified by $\times$ symbols and labeled by the letters $a$ and $b$,
where $a$ is the brighter of the two galaxies (in $g^*$).

\subsection{Completeness}
\label{sec:completness}

The estimate of the completeness of  the merging sample comes from the
differential  distribution of  all SDSS  galaxies in  the survey area.
The  number counts of galaxies in  the five  SDSS bands, $u^*$, $g^*$,
$r^*$, $i^*$, and $z^*$,  in the magnitude range  $12 < r^* < 21$ have
been  studied by \citet{Yasuda01}.   They  found that galaxy counts in
the faintest  magnitude bin  at $r^* >    21$ could be  systematically
affected by galaxies being misclassified as stars.

Figure~\ref{figcomplet}a shows the $g^*$ magnitude distribution of all
the merging  galaxies  (dotted line) and   for all SDSS  EDR  galaxies
(solid  line) in 0.1 magnitude bins  (with  $\sqrt{N}$ error bars).  A
fit is  made to both  samples  --- between  $g^*=16$ and $g^*=20$  for
merging galaxies and between $g^*=16$  and $g^*=21$  for the SDSS  EDR
galaxies;  within  these  limits  the data   are assumed   to be 100\%
complete.  As the magnitude distribution  rolls over, the completeness
is calculated to  be the fraction   of the height of the  distribution
compared to  the fit  representing  100\%  completeness for  the  SDSS
galaxies.     Figure~\ref{figcomplet}b    shows the   $g^*$  magnitude
completeness function.  The merging galaxy sample  is considered to be
nearly 100\% complete in the range $g^* = 16 - 20$.

\subsection{Comparison with a Field Sample}
\label{sec:Comparison}

For each   of the  744 merging  galaxies   with a  known spectroscopic
redshift we chose  the 10 galaxies nearest  in redshift to it from the
full  SDSS EDR spectroscopic galaxy catalog,   without regard to their
RAs and    DECs.   Since  we ignored the     sky  positions of   these
``neighboring'' galaxies, they tend to be  scattered over the full sky
coverage    of  the SDSS EDR.   The    resulting catalog contains 7440
galaxies  with  a  redshift distribution  that   mirrors  that of  the
original 744 merging  galaxies (see Fig~\ref{figRandh}).   This sample
of 7440 galaxies approximates a  sample of field galaxies, since  only
about 10\% of all galaxies lie in rich clusters.

We calculated the  absolute magnitudes and colors  for the galaxies in
each of these two samples --- the sample of  744 merging pair galaxies
with spectroscopic redshifts  and the  sample  of 7440  field galaxies
selected to mimic the former's redshift distribution --- by assuming a
flat cosmological model with $\Omega_{\rm M} = 0.3$, $\Omega_{\Lambda}
=    0.7$, and $H_0   =  100h$~km~s$^{-1}$~Mpc$^{-1}$  and by applying
k-corrections  to the  de-reddened galaxy magnitudes  by  means of the
publicly  available {\tt kcorrect  (v1.10)} code of \citet{Blanton03}.
(Luminosity distances were estimated using the analytical relation of
\citealt{Pen99}.)  As  a  sanity  check,   we plotted   the   absolute
$M_{g^*}-M_{r^*}$ color  vs.\ spectroscopic  redshift for the galaxies
in each  of  these  two samples  (Fig.~\ref{gr_vs_z}).   The  relative
flatness of  E/S0 ridgeline  as  a function   of redshift is   a  good
indicator that our k-corrections are reasonable.

In   Figures~\ref{gmag_vs_color}    and  \ref{rmag_vs_color} we   plot
rest-frame  color-magnitude   diagrams for    these two   samples.  In
Figure~\ref{gmag_vs_color}, we  plot the $g^*$-band absolute magnitude
$M_{g^*}$  vs.\ the rest-frame  colors, since we  selected our merging
pairs  in the $g^*$-band.   In Figure~\ref{rmag_vs_color}, we plot the
$r^*$-band absolute magnitude $M_{r^*}$  vs.\ the rest-frame colors to
simplify comparison with other, $r^*$-band-selected samples (e.g., the
SDSS  EDR catalog of compact  groups  of galaxies by \citealt{Lee03}).
In Figure~\ref{color_vs_color},   we plot  the  rest-frame color-color
diagrams for these two samples.

In Table~\ref{tab:MeanMedianCompare} we list the means and medians for
some properties of the two samples.  We see that the galaxies from the
merging pair sample are, on  average, slightly bluer and slightly less
luminous  than  the galaxies from the   field sample.  One-dimensional
Kolmogorov-Smirnov      (KS)      tests     confirm    these   results
(Fig.~\ref{fig1DKS} and Table~\ref{tab:1DKS}).  This is in contrast to
what  \citet{Lee03} found  for  galaxies in  SDSS EDR compact  groups.
They found  that their sample    of compact group  galaxies were,   on
average,  redder  than galaxies  in their   field  sample.  They  thus
concluded that  their sample of galaxies  in compact group contained a
higher fraction  of Ellipticals than in the  field.  Likewise, here we
conclude  that  our sample of   galaxies in  merging pairs  contains a
higher  fraction of Spirals or  other  star-forming galaxies than does
the field.  (We also note that two-dimensional  KS tests indicate that
the joint  photometric properties of the merging  pair  galaxy and the
field samples differ significantly (Table~\ref{tab:2DKS}).)

These results are  corroborated, albeit  with lower  significance,  by
another morphological classifier.   \citet{Strateva01} found that, for
a   sample   of 147,920 SDSS galaxies     brighter  than $g^*=21$, the
(observer-frame) $u^*-r^*$ color was strongly bimodal.  Early-type (E,
S0, Sa) galaxies mostly lay redward  of the minimum at $u^*-r^*=2.22$,
and  late-type galaxies (Sb,  Sc, Irr)  mostly   lay blueward of  this
limit.  Using this morphological estimator to classify galaxies in our
two redshift samples (Fig.~\ref{fig4}), we
find that only   495$\pm$22   (66.5\%$\pm$3.0\%) of the   744  merging
galaxies are classified  as  early   type compared  with   5223$\pm$72
(70.2\%$\pm$1.0\%) of  the 7440 galaxies  from our field sample (error
bars are based on $\sqrt{N}$ statistics).  

Finally,   we note  that these differences   in  the mean  photometric
properties  of the merging  pair  galaxies and  the field galaxies are
quite  small,  typically only a couple   percent or less.  Although we
cannot completely discount  that these small differences  might merely
be due  to small systematic  offsets between  the flux measurements of
close  pairs and the flux  measurements of field  galaxies,  we do not
believe this  is   the  case.   We  note that   \citet{Infante02} have
visually inspected  the morphologies   and photometry  of  overlapping
galaxies in their  sample, and have   found that these  properties are
correctly determined for pairs of objects  of similar brightness whose
centers are separated by as little as 3~arcsec.  Furthermore, the {\tt
photo} deblender does do a careful  job in conserving the total system
flux   of    blended objects  (\S~4.4.3   of    \citealt{Stoughton02};
\citealt{Lupton01,Lupton02}).   Therefore,  even  though the deblender
might not always apportion the fractional  flux of a blended object to
its individual components with  100\% accuracy (although see the  next
section), the {\em  average\/} flux per  component --- in the case  of
merging pairs, the  total flux for  a given  pair divided by  2 --- is
well determined.  Since the above  analysis concerns the comparison of
the  {\em average} properties of merging   pair galaxies with the {\em
average} properties of field  galaxies,  the results should be  fairly
robust.

\subsection{Holmberg effect}\label{sec:Holmberg}

More  than 40 years ago, \citet{Holmberg58}  discovered that the color
of paired   galaxies were closely  correlated.  This   result has been
interpreted as reflecting a tendency for similar  types of galaxies to
form  and  evolve together.  \cite{Holmberg58}, using  a  sample of 32
galaxy pairs, found that the   (photographic) $B-V$ color indices  for
the individual   galaxies in each   pair showed a   linear correlation
coefficient of   $R=+0.80\pm0.06$;   later   studies   \citep{Tomov78,
Tomov79, Vardanian80,  Demin81, Reshetnikov98} tended to  confirm this
result, finding typical  linear correlation coefficients of $R \approx
0.8$ for the $B-V$ colors of paired galaxies (see, e.g., the review by
\citealt{Karachentsev87}).   None of these studies, however, contained
many more than 100 galaxy pairs.

Here, we  calculate  the Holmberg Effect   in the SDSS colors  for our
sample   of  1479   merging    pairs.    We  plot   our    results  in
Figure~\ref{fig5}.  (A clickable version of  the $g^*-r^*$ plot, which
allows the user to see the corresponding SDSS  image for each pair, is
available  at our previously  mentioned  public website.)  Calculating
the linear correlation coefficients, we find
\begin{eqnarray}
R_{u^*-g^*} & = & 0.20 \pm 0.03  \\
R_{g^*-r^*} & = & 0.38 \pm 0.03  \nonumber \\
R_{r^*-i^*} & = & 0.19 \pm 0.03  \nonumber \\
R_{i^*-z^*} & = & 0.07 \pm 0.04  \nonumber 
\end{eqnarray}
where   the (1$\sigma$)  errors  were  calculated  via  1000 bootstrap
resamplings of the original data.

We note that  the level and   the significance of the the  correlation
depends  on the  color index.   It  is  highest and has  the  greatest
significant ($> 12\sigma$) for  $g^*-r^*$,  which is the closest  SDSS
analog to $B-V$.    It is lowest  and has  the least significance  ($<
2\sigma$) for $i^*-z^*$, which is  not surprising, since most galaxies
have very similar  $i^*-z^*$  colors \citep{Fukugita95} and thus  even
small random  photometric errors ($<$ 0.05~mag)  will tend to wash out
any correlations.

Although  apparently  statistically significant,  could these measured
correlations be  due to systematic  errors  in our measurements rather
than  due   to a  physical   effect?   The  most   obvious  suspect is
cross-contamination in the galaxy photometry.  After  all, it is quite
possible that the colors of two galaxies in a close pair merely appear
similar because the photometry of galaxy  ``a'' is contaminated by the
photometry of  galaxy  ``b''.  To test, we   plotted the difference in
colors between  the two  galaxies  in  each  pair vs.   their  angular
separation  on the   sky (Fig.~\ref{holmbergtest}).   Although  not  a
conclusive  test, one  would expect ---   if the Holmberg  effect were
entirely  due to such systematic    cross-contamination in the  galaxy
photometry ---  that more widely separated  pairs would show less of a
color  concordance  than less widely  separated  pairs.  We do not see
such  a  trend in  our   data (Fig.~\ref{holmbergtest}).   

We note that  our estimate for $R_{g^*-r^*}$  is only  about half that
measured for $R_{B-V}$ in previous studies.  This is surprising, since
$g^*-r^* \approx    1.0  \times  (B-V) -   0.2$    for normal  stellar
populations (see  Table~7 of \citealt{Smith02}).   Therefore, a  color
concordance  in $g^*-r^*$ should  be roughly  as strong  as the one in
$B-V$.  Two  possible explanations come  immediately  to mind.  First,
since  we have   limited spectroscopic redshift   information  for our
sample, it  may  suffer  from contamination by   spurious  pairs whose
members have substantially discordant redshifts.  Since spurious pairs
should  show no  color   concordance   (there  would be  no   physical
connection between the  members  of  such pairs), this   contamination
would tend to weaken the appearance of any underlying Holmberg effect.
We  do not think, however, that  our sample is  unduly contaminated by
spurious pairs.  For  those 40  pairs in  our catalog for  which we do
have spectroscopic information for both galaxies, only 5 --- or 12.5\%
--- show obviously  discordant redshifts (velocity differences between
the members  of $>$  1000~km~s$^{-1}$).   

A second possible  explanation for the  difference between our measure
of $R_{g^*-r^*}$ and previous measurements of $R_{B-V}$  may be due to
the samples employed.  Our sample is  a sample of {\em merging} pairs,
which have  typical projected  intra-pair separations  of less  than a
galaxy diameter.  The samples   used in previous studies   (e.g., from
\citealt{VV59}  and    \citealt{Karachentsev72})   were  typically  of
interacting but not necessarily merging pairs and have on average much
wider intra-pair  separations.  This admittedly  is counter-intuitive:
one would   expect that merging  pairs  would have  a greater level of
color concordance than do their more widely separated cousins.  On the
other   hand,  it  is during  these  close  encounters  when most mass
transfers and starburst triggers occur.  It  is precisely at this time
of chaotic inequilibrium when  the two members  of  a pair may  appear
quite different from each other. 

This latter explanation is supported by results from \citet{Balogh03},
who studied the environmental dependence of star formation rate in the
local Universe using data from both the 2dF Galaxy Redshift Survey and
the SDSS,  and \citet{Lambas02}, who  studied  the star formation rate
for  isolated galaxy pairs  in  the 100~K  public  release of  the 2dF
Galaxy Redshift Survey.   \citet{Balogh03} note that the  {\em only\/}
environment they  found  where  significantly  enhanced   $H_{\alpha}$
emission is generally  observed is  for  very close galaxy pairs  with
projected separations    of  less  than $35h^{-1}$~kpc   and  velocity
differences  less  than  150~km/s; wider  pairs do  not  show  such an
enhancement.  Similarly, \citet{Lambas02} find  that star formation is
significantly  enhanced  only for   very close  pairs ---  those  with
projected  separations   less    than  $25h^{-1}$~kpc  and    velocity
differences less than 100~km/s.


\section{Conclusion}
\label{sec:conclusion}

We have  presented an  algorithm  for the  automated identification of
merging  galaxy pair candidates  from  the SDSS data.  The  algorithm,
which implements a variation of \citet{Karachentsev72}'s isolated pair
criteria, proves to be very efficient and fast.  

Within the $\approx$462~sq~deg of the  SDSS EDR photometric data  set,
we have identified 1479 merging galaxy pair candidates for galaxies in
the magnitude range $16.0 \leq g^* \leq  21.0$, where we define {\em a
merging pair\/} as  one in which the two  galaxy centers are separated
by  less than   the  sum  of the  members'  Petrosian radii.   By  our
definition, we estimate that approximately 0.5\%  of SDSS EDR galaxies
are members   of merging pairs.   We  provide a  catalog and an online
atlas of all 1479 merging pair candidates.

Analysis indicates that the merging pair  galaxies tend to be slightly
bluer     than a  corresponding  field    sample   from the SDSS  EDR.
Furthermore, compared to the field sample,  the merging pair sample is
weighted toward  later type galaxies.   We take this  result of active
star  formation occurring within merging pair  galaxies.  

Finally, we studied the  color concordance of  galaxy members  in each
pair.  Using much smaller samples,  \citet{Holmberg58} and others have
found the linear correlation coefficient for  the $B-V$ colors of pair
galaxies to be $R_{B-V}  \approx 0.8$.  We  find that the strength  of
the color concordance varies  drastically with  color index, with  the
$g^*-r^*$ colors of   pair  galaxies showing  the   strongest and most
significant  concordance ($R_{g^*-r^*}=0.38\pm0.03$) and the $i^*-z^*$
colors  showing    the   weakest and  least   significant  concordance
($R_{i^*-z^*}=0.07\pm0.04$).  Surprisingly, although  $g^*-r^* \approx
B-V$  for stellar  populations, we  find   that  the $g^*-r^*$   color
concordance  we measure is  only about half the  strength of the $B-V$
color  concordance  measured in  other samples.    We expect that  the
difference  is due to sample properties:  the galaxies in  an SDSS EDR
merging pair are on average much closer  together than the galaxies in
a pair from one  of these other samples.   It may be that the galaxies
in SDSS  EDR merging pairs  are  undergoing more vigorous and  chaotic
star formation compared with the galaxies in the generally wider pairs
characteristic of these other samples.


\acknowledgments    

The authors wish to express their  thanks to the anonymous referee for
comments that improved the presentation of this material.

Funding for the creation and distribution of the SDSS Archive has been
provided    by  the Alfred   P.  Sloan   Foundation, the Participating
Institutions, the  National Aeronautics and  Space Administration, the
National Science   Foundation,  the U.S.   Department of  Energy,  the
Japanese Monbukagakusho, and the Max Planck Society. The SDSS Web site
is {\tt http://www.sdss.org/}.

The SDSS is managed by the Astrophysical Research Consortium (ARC) for
the Participating Institutions. The Participating Institutions are The
University of Chicago, Fermilab, the Institute for Advanced Study, the
Japan Participation Group, The   Johns Hopkins University,  Los Alamos
National Laboratory,  the Max-Planck-Institute  for  Astronomy (MPIA),
the  Max-Planck-Institute for  Astrophysics  (MPA), New   Mexico State
University, University of Pittsburgh, Princeton University, the United
States Naval Observatory, and the University of Washington.

This research has made   use of the NASA/IPAC  Extragalactic  Database
(NED),  which is operated by the  Jet  Propulsion Laboratory, Caltech,
under contract with the National Aeronautics and Space Administration.



\clearpage

\begin{deluxetable}{rrrrrr}
\tabletypesize{\small}
\tablecaption{Percentage of merging galaxies in the SDSS EDR ($g^*=16-21$).\label{tabinmerg}}
\tablewidth{0pt}
\tablenum{1}
\tablehead{
\colhead{Run} & 
\colhead{area} & 
\colhead{$N_{\rm gal}^{\rm tot}$} & 
\colhead{$N_{\rm gal}^{\rm pair}$} &  
\colhead{$\Sigma_{\rm gal}^{\rm pair}$} &  
\colhead{$f_{\rm gal}^{\rm pair}$} \\ 
\colhead{} & 
\colhead{[$deg^{2}$]} & 
\colhead{} & 
\colhead{} &  
\colhead{[\# deg$^{-2}$]} &  
\colhead{}  
}
\startdata
  94  & 83  &119107 &  590 & 7.1  & 0.0050 \\
 125  & 83  &117235 &  378 & 4.55 & 0.0032 \\
 752  & 114 &142785 &  716 & 4.53 & 0.0050 \\
 756  & 114 &141124 &  958 & 8.4  & 0.0068 \\
1336  & 16  & 20217 &   92 & 5.75 & 0.0046 \\
1339  & 16  & 19379 &   84 & 5.25 & 0.0043 \\
1356  & 17  & 20191 &   80 & 4.7  & 0.0040 \\
1359  & 19  & 19610 &   60 & 3.15 & 0.0031 \\
total & 462 &458524 & 2958 & 6.4  & 0.0065 \\
\enddata
\end{deluxetable}

\clearpage

{\tiny
\begin{deluxetable}{c@{\,}c@{\,}c@{\,}c@{\,}c@{\,}c@{\,}c@{\,}c@{\,}c@{\,}c@{\,}c@{\,}c@{\,}c@{\,}c}
\tabletypesize{\tiny}
\tablecaption{Properties of the SDSS EDR Merging Pairs Sample.\label{tabmerg}}
\tablewidth{0pt}
\tablenum{2}
\tablecolumns{9}
\tablehead{
\colhead{ID}  & 
\colhead{RA}  & 
\colhead{DEC}  & 
\colhead{$u^*$}& 
\colhead{$g^*$}& 
\colhead{$r^*$}& 
\colhead{$i^*$}& 
\colhead{$z^*$}& 
\colhead{Sep}& 
\colhead{A}& 
\colhead{$z_A$}& 
\colhead{B}& 
\colhead{$z_B$}  \\
 \colhead{} &
 \colhead{[J2000.0]} &
 \colhead{[J2000.0]} &
 \colhead{} &
 \colhead{} &
 \colhead{} &
 \colhead{} &
 \colhead{} &
 \colhead{[arcsec]} &
 \colhead{} &
 \colhead{} & 
 \colhead{} & 
 \colhead{} & 
 \colhead{} \\ 
 \colhead{(1)} &
 \colhead{(2)} &
 \colhead{(3)} &
 \colhead{(4)} &
 \colhead{(5)} &
 \colhead{(6)} &
 \colhead{(7)} &
 \colhead{(8)} &
 \colhead{(9)} &
 \colhead{(10)} &
 \colhead{(11)} & 
 \colhead{(12)} & 
 \colhead{(13)}  \\ 
 }
\startdata
1& 00:02:49.25& +00:45:05.77& 18.544& 16.367& 15.369& 14.818& 14.463& 5.856& SDSS J000249.06+004504.83 &0.086497 &SDSS J000249.43+004506.72& - \\
2& 00:03:57.29& +00:28:29.21& 19.137& 17.788& 16.944& 16.509& 16.193& 5.708& SDSS J000357.27+002832.06 &0.084922 &SDSS J000357.30+002826.37& - \\
3& 00:06:46.67& -00:36:55.26& 20.487& 19.401& 18.725& 18.411& 18.128& 5.633& SDSS J000646.74-003652.62 &-  &SDSS J000646.60-003657.90& - \\
4& 00:07:17.63& +00:50:20.14& 18.257& 16.706& 15.912& 15.491& 15.143& 12.453& SDSS J000717.23+005018.52 &0.102078 &SDSS J000718.03+005021.76& - \\
5& 00:07:20.49& +00:29:34.40& 18.113& 16.128& 15.122& 14.666& 14.309& 8.690& SDSS J000720.21+002933.20 &-  &SDSS J000720.77+002935.60& 0.091190\\
6& 00:07:55.69& +00:32:55.12& 19.589& 18.458& 17.962& 17.659& 17.540& 5.530& SDSS J000755.59+003252.81 &-  &SDSS J000755.79+003257.44& - \\
7& 00:08:26.77& +00:22:13.31& 19.042& 17.578& 16.918& 16.458& 16.076& 11.506& SDSS J000826.52+002209.00 &0.054646 &SDSS J000827.02+002217.61& - \\
8& 00:08:32.70& +01:02:20.20& 18.806& 17.235& 16.564& 16.083& 15.824& 4.089& SDSS J000832.83+010219.86 &0.060147 &SDSS J000832.56+010220.54& - \\
\enddata
\end{deluxetable}
}

\clearpage

\begin{deluxetable}{lrrrrr}
\tabletypesize{\small}
\tablecaption{Mean and Median Properties of the SDSS EDR Merging Pairs Sample. \label{tabin}}
\tablewidth{0pt}
\tablenum{3}
\tablehead{
\colhead{Merging pairs}& 
\colhead{mean}& 
\colhead{median}& 
\colhead{mean\tablenotemark{\dag}}& 
\colhead{median\tablenotemark{\dag}}& 
\colhead{}  
}
\startdata
 $u^*$ &19.37$\pm$0.027 & 19.43 & 19.18$\pm$0.027 & 19.25 &\\
 $g^*$ &17.91$\pm$0.026 & 18.00 & 17.77$\pm$0.026 & 17.86 &\\
 $r^*$ &17.11$\pm$0.026 & 17.18 & 17.01$\pm$0.026 & 17.07 &\\
 $i^*$ &16.73$\pm$0.026 & 16.77 & 16.65$\pm$0.026 & 16.69 &\\
 $z^*$ &16.47$\pm$0.028 & 16.46 & 16.41$\pm$0.028 & 16.41 &\\
Projected Separation [arcsec] &  6.12$\pm$0.086 &  5.24  & & & \\ 
$u^*_{a}$-$u^*_{b}$ &1.18$\pm$0.031 & 1.01& 1.17$\pm$0.031 & 1.01& \\ 
$g^*_{a}$-$g^*_{b}$ &1.29$\pm$0.026 & 1.06& 1.28$\pm$0.026 & 1.06\\ 
$r^*_{a}$-$r^*_{b}$ &1.32$\pm$0.029 & 1.12& 1.32$\pm$0.029 & 1.12&\\ 
$i^*_{a}$-$i^*_{b}$ &1.32$\pm$0.030 & 1.14& 1.31$\pm$0.030 & 1.13&\\ 
$z^*_{a}$-$z^*_{b}$ &1.36$\pm$0.033 & 1.19& 1.35$\pm$0.033 & 1.19&\\ 
\enddata
\tablenotetext{\dag}{After applying the reddening correction using the \citet{Schlegel98} maps. }
\end{deluxetable}

\clearpage

\begin{deluxetable}{lrrrr}
\tabletypesize{\small}
\tablecaption{Means and Medians from Merging Galaxy and Field Galaxy Redshift Samples.
\label{tab:MeanMedianCompare}}
\tablewidth{0pt}
\tablenum{4}
\tablehead{
\colhead{Test}  & 
\multicolumn{2}{c}{Merging Pair Galaxies (744)}  & 
\multicolumn{2}{c}{Field Galaxies (7440)} \\
\colhead{}  & 
\colhead{mean}& 
\colhead{median}& 
\colhead{mean}& 
\colhead{median}
}
\startdata
$z_{\rm sp}$	 &   0.1107$\pm$0.0019&   0.104&   0.1107$\pm$0.0006&  0.104\\
$M(g^*)$         & -20.134$\pm$0.048&   -20.337& -20.243$\pm$0.015& -20.367\\ 
$M(r^*)$         & -20.767$\pm$0.052&   -20.986& -20.898$\pm$0.016& -21.047\\ 
$M(u^*)-M(g^*)$  &   1.406$\pm$0.015&     1.534&   1.444$\pm$0.004&   1.534\\   	
$M(g^*)-M(r^*)$  &   0.632$\pm$0.008&     0.714&   0.658$\pm$0.002&   0.709\\ 
$M(r^*)-M(i^*)$  &   0.315$\pm$0.007&     0.368&   0.343$\pm$0.001&   0.370\\ 	
$M(i^*)-M(z^*)$	 &   0.246$\pm$0.010&     0.264&   0.262$\pm$0.002&   0.271\\
\enddata
\end{deluxetable}

\clearpage

\begin{deluxetable}{lcc}
\tabletypesize{\small}
\tablecaption{Results from 1D Kolmogorov-Smirnov Tests. \label{tab:1DKS}}
\tablewidth{0pt}
\tablenum{5}
\tablehead{
\colhead{Test}& 
\colhead{KS Statistic}& 
\colhead{Probability}
}
\startdata
$z_{\rm sp}$     & 0.0019  &1.00000\\
$M(g^*)$         & 0.0409  &0.20879\\
$M(r^*)$         & 0.0472  &0.09850\\
$M(u^*)-M(g^*)$  & 0.0812  &0.00027\\ 
$M(g^*)-M(r^*)$  & 0.0796  &0.00038\\ 
$M(r^*)-M(i^*)$  & 0.0933  &0.00002\\
$M(i^*)-M(z^*)$  & 0.0659  &0.00566\\
\enddata
\end{deluxetable}

\clearpage

\begin{deluxetable}{lcc}
\tabletypesize{\small}
\tablecaption{Results from 2D Kolmogorov-Smirnov Tests \label{tab:2DKS}}
\tablewidth{0pt}
\tablenum{6}
\tablehead{
\colhead{Test}  & 
\colhead{KS Statistic} &
\colhead{Probability} 
}
\startdata
$z_{\rm sp}$        vs.\ $M(g^{*})-M(r^{*})$   &0.081 &0.00520\\
                                               &      &      \\
$M(u^{*})-M(g^{*})$ vs.\ $M(g^{*})-5\log h$    &0.090 &0.00051\\
$M(g^{*})-M(r^{*})$ vs.\ $M(g^{*})-5\log h$    &0.094 &0.00020\\
$M(r^{*})-M(i^{*})$ vs.\ $M(g^{*})-5\log h$    &0.095 &0.00034\\
$M(i^{*})-M(z^{*})$ vs.\ $M(g^{*})-5\log h$    &0.083 &0.00412\\
                                               &      &      \\
$M(u^{*})-M(g^{*})$ vs.\ $M(r^{*})-5\log h$    &0.097 &0.00039\\
$M(g^{*})-M(r^{*})$ vs.\ $M(r^{*})-5\log h$    &0.100 &0.00017\\
$M(r^{*})-M(i^{*})$ vs.\ $M(r^{*})-5\log h$    &0.093 &0.00064\\
$M(i^{*})-M(z^{*})$ vs.\ $M(r^{*})-5\log h$    &0.088 &0.00019\\
                                               &      &      \\
$M(g^{*})-M(r^{*})$ vs.\ $M(u^{*})-M(g^{*})$   &0.118 &$1.4\times10^{-6}$\\
$M(r^{*})-M(i^{*})$ vs.\ $M(g^{*})-M(r^{*})$   &0.147 &$1.2\times10^{-9}$\\
$M(i^{*})-M(z^{*})$ vs.\ $M(r^{*})-M(i^{*})$   &0.114 &$1.3\times10^{-5}$\\
\enddata
\end{deluxetable}


\clearpage

\clearpage
\begin{figure}
 \centering
\caption{Polar view of SDSS EDR spectroscopic sample (black symbols) and the sample 
of merging galaxies with redshift (red symbols).  The radial coordinate is redshift;
the angular coordinate is RA. \label{fig1}}
\end{figure}  

\clearpage
\begin{figure}
\centering
\makebox[85mm]{\psfig{file=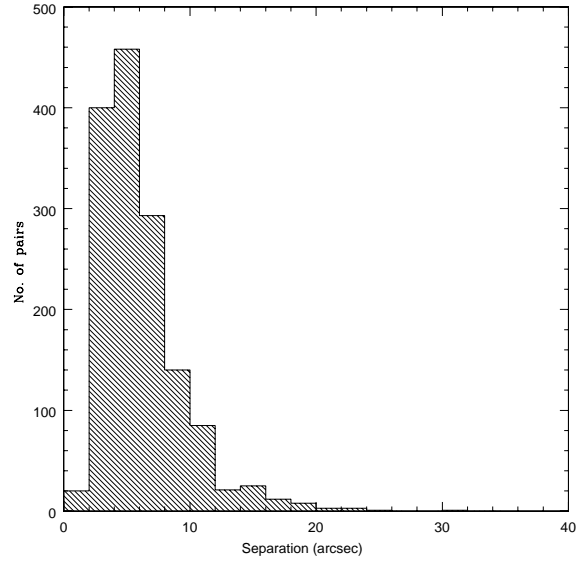,width=80mm,height=80mm,clip=,angle=0,silent=}}
\vspace{-.25cm}
\caption{Histogram of the projected separations in arcsec for the SDSS EDR merging pairs. \label{projsephist}}
\end{figure}

\clearpage
\begin{figure}
\centering
\makebox[125mm]{\psfig{file=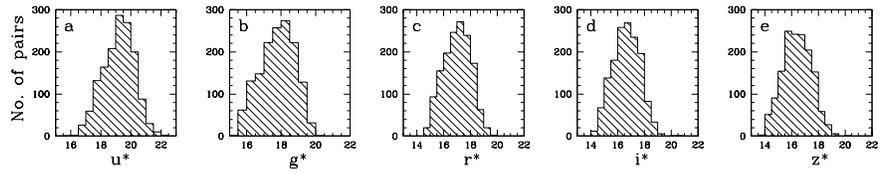,width=120mm,height=120mm,clip=,angle=0,silent=}}
\vspace{-9.5cm}
\caption{Histogram of the magnitudes for the SDSS EDR merging pair galaxies 
in each of the SDSS filters. \label{figh5bandmag}}
\end{figure}

\clearpage
\begin{figure}
\centering
\makebox[125mm]{\psfig{file=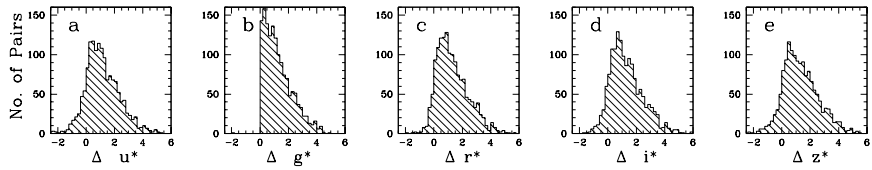,width=120mm,height=120mm,clip=,angle=0,silent=}}
\vspace{-9.5cm}
\caption{Histogram of the  difference in magnitude between  the members in each
pair for the SDSS EDR merging pairs in each of the 5 SDSS filters. The
clear sharp   cut  in $g^{*}$   is  due  to  the   sample  selection 
(see \S~\ref{sec:SelectionCriteria}). \label{fighindifmag}}
\end{figure}

\clearpage
\begin{figure}
\centering
\makebox[120mm]{\psfig{file=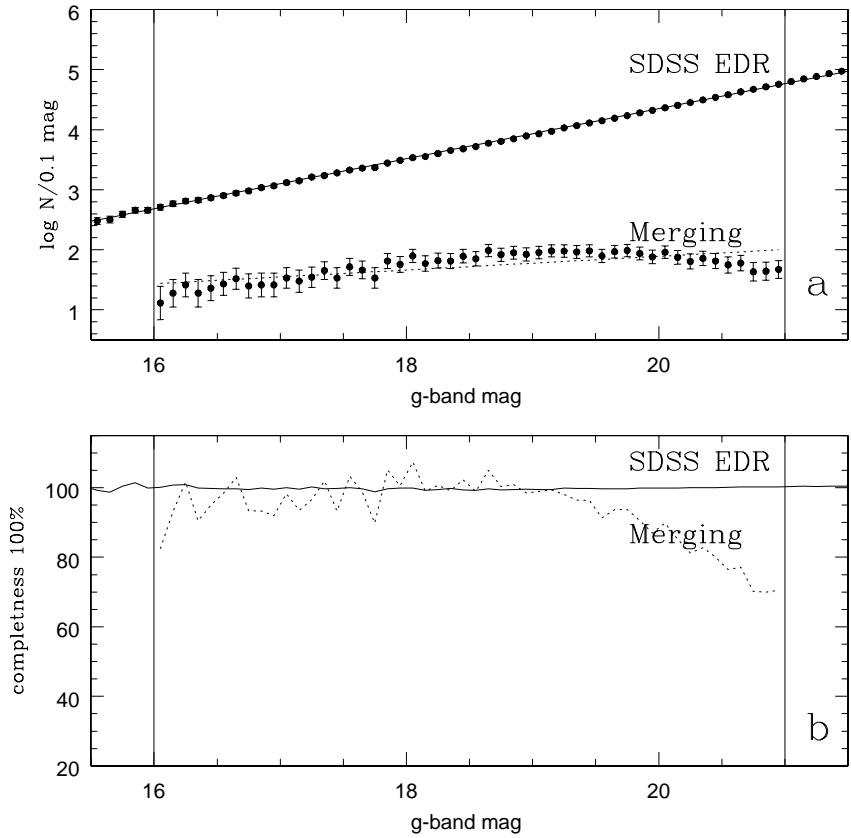,width=120mm,height=120mm,clip=,angle=0,silent=}}
\caption{(a) The $g^*$ magnitude distribution in  0.1  magnitude bins of all SDSS
EDR galaxies and of SDSS EDR merging pair galaxies.  The thin straight
lines are fits to these distributions (solid line for SDSS EDR, dotted
line  for merging galaxies).   (b) The  $g^*$-band completeness of all
SDSS EDR galaxies (solid line)  and of SDSS  EDR merging pair galaxies
(dotted line).
\label{figcomplet}}
\end{figure}  

\clearpage
\begin{figure}
\centering
\makebox[85mm]{\psfig{file=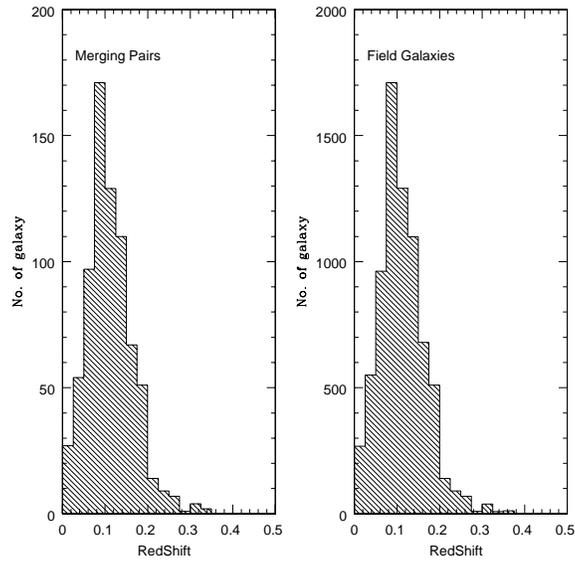,width=80mm,height=80mm,clip=,angle=0,silent=}} 
\caption{Histogram of the (a) SDSS EDR merging pair galaxies with redshift, 
(b) the field sample. \label{figRandh}}
\end{figure}

\clearpage
\begin{figure*}
\centering
\caption{De-reddened and k-corrected $M_{g^*}-M_{r^*}$ color vs.\ 
spectroscopic redshift. {\em Black  dots:} field galaxy  sample.  {Red
dots:} merging  galaxy sample.  (Note:  Merging galaxy sample only  contains
those 744 galaxies with known spectroscopic redshifts.)
\label{gr_vs_z}}
\end{figure*}

\clearpage
\begin{figure*}
\centering
\caption{The SDSS $g^*$ band absolute magnitude vs.\ color (de-reddened 
\& k-corrected). {\em Black dots:} field galaxy  sample. {Red dots:} Merging 
galaxy sample.(Note:  Merging galaxy  sample  only contains  those 744
galaxies with known spectroscopic redshifts.) 
\label{gmag_vs_color}}
\end{figure*}

\clearpage
\begin{figure*}
\centering
\caption{The SDSS $r^*$ band absolute magnitude vs.\ color (de-reddened 
\& k-corrected). {\em Black dots:}  field galaxy sample. {Red dots:} Merging 
galaxy sample.  (Note:  Merging galaxy sample only contains those 744 
galaxies with known spectroscopic redshifts.)
\label{rmag_vs_color}}
\end{figure*}

\clearpage
\begin{figure*}
\centering
\caption{Color vs.\ Color (de-reddened \& k-corrected)
{\em Black dots:}  field galaxy sample. {Red dots:} Merging galaxy sample.  
(Note:  Merging galaxy sample only contains those 744 galaxies with 
known spectroscopic redshifts.)
\label{color_vs_color}}
\end{figure*}

\clearpage
\begin{figure*}
\centering
\makebox[180mm]{\psfig{file=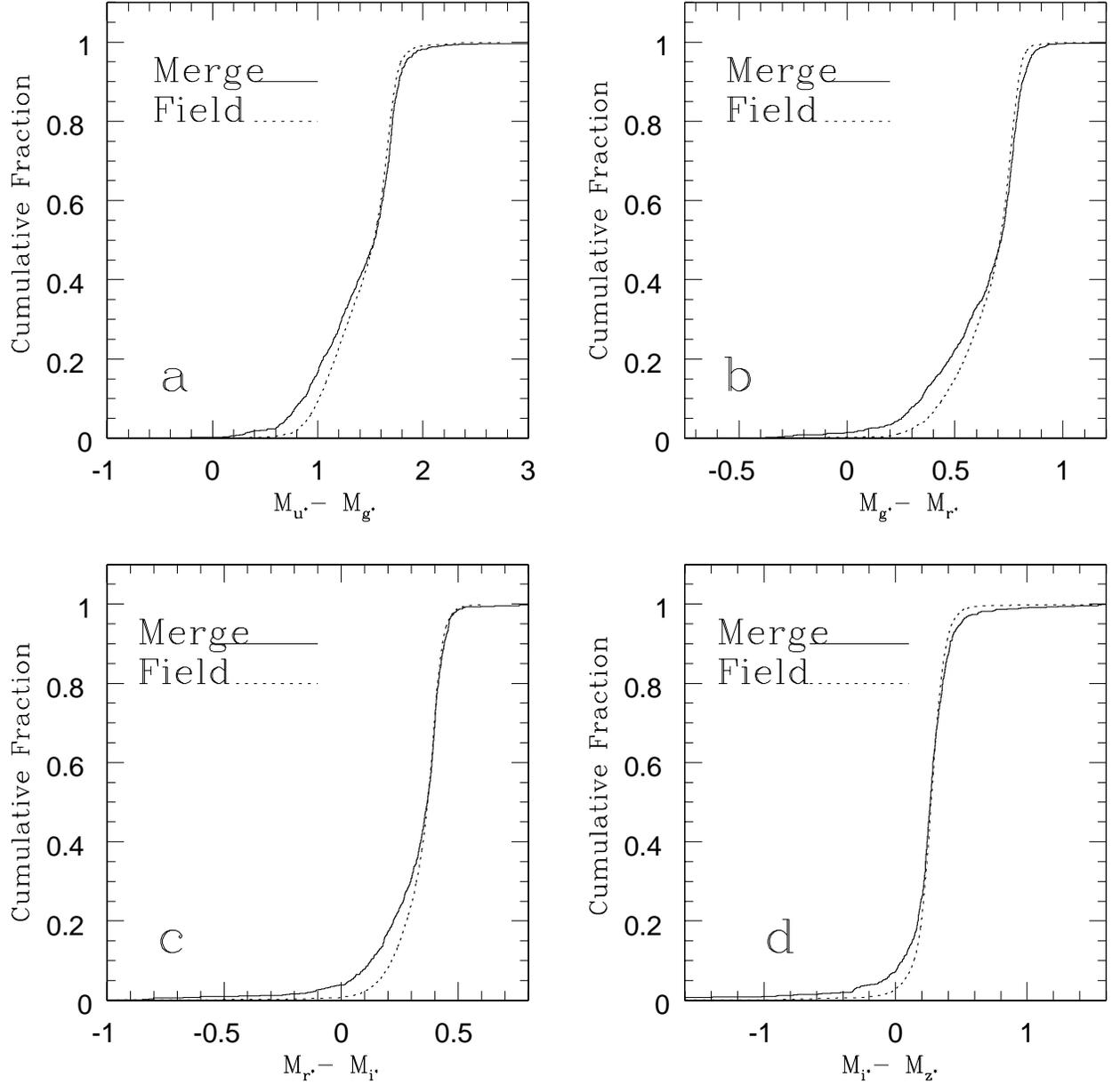,width=178mm,height=178mm,rwidth=175mm,clip=,angle=0,silent=}}
\caption{One-dimensional KS tests comparing the sample of 744 merging pair 
galaxies with spectroscopic redshifts and the 7440 field galaxy comparison
sample.
\label{fig1DKS}}
\end{figure*}

\clearpage
\begin{figure}
\centering
\caption{(a) $u^*-r^*$ vs. $g^*$ color-magnitude diagram for the sample 
of 744 merging galaxies   with redshifts  (red  symbols) and  for  the
sample  of 7440 galaxies  in the field   sample.  The vertical line at
$u^*-r^* = 2.22$ represents the line of demarcation between early- (E,
S0, Sa) and late-type (Sb, Sc, Irr) galaxies as determined by
\citet{Strateva01}.  The slope at the faint limit in $g^*$ is due
to the fact that the main spectroscopic galaxy sample for the SDSS EDR
was selected in $r^*$, not in $g^*$ \citep{Strauss02}.
\label{fig4}}
\end{figure}  

\clearpage
\begin{figure}
\centering
\caption{Study of the Holmberg effect for SDSS EDR merging pair galaxies.
Galaxies in Elliptical-Elliptical pairs are denoted by red symbols, galaxies
in Spiral-Elliptical pairs by green symbols, galaxies in Elliptical-Spiral
pairs by black symbols, and galaxies in Spiral-Spiral pairs by blue symbols.
Galaxy types  were determined using the \citet{Strateva01} $u^*-r^*$ 
morphological classifier.   \label{fig5}}
\end{figure}  

\clearpage
\begin{figure}
\centering
\makebox[90mm]{\psfig{file=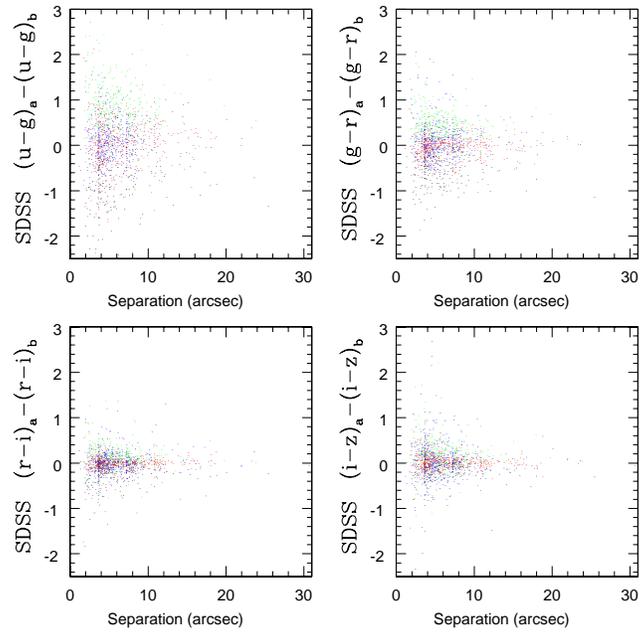,width=90mm,height=90mm,clip=,angle=0,silent=}}
\caption{Test of the Holmberg effect against intra-pair separation for SDSS EDR merging pair galaxies.
The color coding of the symbols is the same as in Fig.~\ref{fig5}.
\label{holmbergtest}}
\end{figure}  

\end{document}